\documentclass[english,aps,floats,twocolumn,showpacs,nofootinbib]{revtex4}
\usepackage{pslatex}
\usepackage[T1]{fontenc}
\usepackage[latin1]{inputenc}
\usepackage{graphicx}
\usepackage{epsfig} 

\usepackage{calc}
\usepackage{ifthen}

{
{
{
\newcommand{\bea}{\begin{eqnarray}}
\newcommand{\eea}{\end{eqnarray}}

\newcommand{\nc}{\newcommand}
\nc{\renc}{\renewcommand}
\nc{\eqs}[2]{\mbox{Eqs.~(\ref{#1},\,\ref{#2})}}
\nc{\eq}[1]{\mbox{Eq.~(\ref{#1})}}
\nc{\figs}[2]{\mbox{Figs.~(\ref{#1},\,\ref{#2})}}
\nc{\fig}[1]{\mbox{Fig~.(\ref{#1})}}
\nc{\be}[1]{\begin{equation} \mbox{$\label{#1}$}}
\nc{\ee}{\vspace{0.1cm}\end{equation}}

\newcommand{\bean}{\begin{eqnarray*}}
\newcommand{\eean}{\end{eqnarray*}}

\def\GeV{{\rm \ GeV}}

\def\TeV{{\rm \ TeV}}

\def\lae{\;^{<}_{\sim} \;} \def\gae{\; ^{>}_{\sim} \;}

\def\oM{\overline{M}}

\def\cP{{\cal P}}

\begin{document}
\title{Anthropically Selected Baryon Number and Isocurvature Constraints}
\author{John McDonald}
\email{j.mcdonald@lancaster.ac.uk}
\affiliation{Lancaster-Manchester-Sheffield Consortium for Fundamental Physics, Cosmology and Astroparticle Physics Group, Dept. of Physics, University of 
Lancaster, Lancaster LA1 4YB, UK}
\begin{abstract}

     The similarity of the observed baryon and dark matter densities suggests that they are physically related,  either via a particle physics mechanism or anthropic selection. A pre-requisite for anthropic selection is the generation of superhorizon-sized domains of different $\Omega_{B}/\Omega_{DM}$. Here we consider generation of  domains of different baryon density via random variations of the phase or magnitude of a complex field $\Phi$ during inflation. Baryon isocurvature perturbations are a natural consequence of any such mechanism. We derive baryon isocurvature bounds on the expansion rate during inflation $H_{I}$ and on the mass parameter $\mu$ which breaks the global $U(1)$ symmetry of the $\Phi$ potential. We show that when $\mu \lae H_{I}$ (as expected in SUSY models) the baryon isocurvature constraints can be satisfied only if $H_{I}$ is unusually small, $H_{I} < 10^{7} \GeV$, or if non-renormalizable Planck-suppressed corrections to the $\Phi$ potential are excluded to a high order. Alternatively, an unsuppressed $\Phi$ potential is possible if $\mu$ is sufficiently large, $\mu \gae 10^{16} \GeV$. We show that the baryon isocurvature constraints can be naturally satisfied in Affleck-Dine baryogenesis, as a result of the high-order suppression of non-renormalizable terms along MSSM flat directions. 

\end{abstract}
\maketitle

\section{Introduction}

      The cosmological dark matter and baryon mass densities are observed to be within an order of magnitude of each other, $\Omega_{B}/\Omega_{DM} \approx 1/5$ \cite{wmap}. However, baryogenesis and dark matter production are often physically unrelated in particle physics models. 
So why are these densities similar? 

     It is possible to produce both dark matter and baryon number simultaneously, thereby directly relating their number densities. Such models are usually based on an overall conserved charge which is shared by the baryons and dark matter particles. This implies that dark matter is asymmetric with a small dark matter particle mass, $m_{DM} \sim 1-10 \GeV$.

      However, in the case where thermal relic WIMPs are the explanation for dark matter, a particle physics mechanism cannot simply relate the baryon and dark matter number densities directly to each other (as in the charge conservation models), but must specifically relate the baryon asymmetry  to the thermal relic WIMP density. This implies a connection between the weak annihilation freeze-out process responsible for the thermal relic WIMP density and the mechanism determining the observed baryon asymmetry. Recently there have been some proposals which make this connection, based either on the modification of a pre-existing baryon asymmetry ({\it baryomorphosis}) \cite{bm,bm2} or on the generation of the baryon asymmetry via annihilation of dark matter ({\it WIMPy baryogenesis}) \cite{wimpy}. Such mechanisms require a number of additional particles and are strongly constrained by B washout. Since the new particles are necessarily at the TeV scale, these models may be testable at the LHC. 

    The alternative is anthropic selection. Anthropic selection models have two components: (i) a mechanism to generate domains\footnote{By 'domain' we mean any patch of the Universe with different conditions from ours. This could also include domains in different inflationary patches, although we will focus on a single inflated patch.} with varying $\Omega_{B}/\Omega_{DM}$ and (ii) the assumption that domains with $\Omega_{B}/\Omega_{DM} \sim 1$ are favoured by the evolution of observers. 
An example of such a model was proposed in \cite{lindeax}. 
In this model dark matter is due to a condensate of axions with a domain-dependent density, while the baryon number density is assumed fixed. Domains with average dark matter densities larger than in our domain result in the formation of galaxies with baryon and dark matter densities which are strongly enhanced relative to the average. The enhancement is due to perturbations becoming non-linear earlier \cite{lindeax}. The enhanced dark matter and baryon densities in galaxies are then assumed to provide the required anthropic cut-off. 

   However, if the dark matter density is fixed throughout the Universe, as in the case of thermal relic WIMPs, we need an alternative way to vary $\Omega_{B}/\Omega_{DM}$. Here we consider varying the baryon density between domains. It is, in principle, easy to vary the baryon density on superhorizon scales. All that is necessary is that the CP-violating phase or strength of B-violation depends on a field which is effectively massless until the onset of baryogenesis. During inflation the field can take random values on scales much larger than the horizon when the observed Universe exits the horizon at $N = 60$ e-foldings before the end of inflation. Therefore superhorizon domains with different baryon number will exist at present. It is therefore likely that there will exist some domains with $\Omega_{DM} \sim \Omega_{B}$. It is also likely that the largest field value (magnitude or phase) will have the largest probability, in which case we will most likely live in a domain with the largest possible baryon asymmetry up to anthropic selection effects. As in the axion model, an average baryon density which is larger than the observed baryon density will be enhanced to a much larger baryon density in galaxies, which may then serve as an anthropic cut-off. We will refer to such models as {\it anthropic baryogenesis} models in the following.

       In order for the baryon density in a domain to be random, it should not be determined purely by the parameters of the $\Phi$ potential. For example, suppose the CP-violating phase $\theta$ of a complex field $\Phi = \phi e^{i \theta}/\sqrt{2}$, which is effectively massless during inflation ($m_{\Phi}^{2} \ll H_{I}^{2}$), determines the baryon asymmetry. (The CP-conserving direction can be defined to be $\theta = 0$.) In a 'typical' domain we expect $\theta \sim \pi$. The baryon asymmetry will then be near maximal and will be essentially determined by the parameters of the $\Phi$ potential. As a result, a coincidence between the maximal baryon asymmetry and the DM density is required; there is no real anthropic selection. In order to have a randomly-varying baryon asymmetry, the domain which has the observed baryon asymmetry must be {\it atypical}, with $\theta \ll 1$. In this case there is no direct connection between the baryon asymmetry and the parameters of the potential and so no element of coincidence. Moreover, the baryon density in neighbouring domains can then be much larger or smaller than in our domain, allowing anthropic selection to function, whereas in the case where $\theta \sim \pi$ only O(1) fractional increases in the baryon asymmetry relative to our domain are possible.   

      However, the dependence on a massless complex scalar has a consequence that will impose a strong constraint on any anthropic selection mechanism of this type; quantum fluctuations of the massless field will produce baryon isocurvature pertubations. We will show that the atypically small value of $\theta$ enhances the baryon isocurvature perturbations, resulting in strong constraints on anthropic baryogenesis models. 

        The paper is organized as follows. In Section 2 we discuss the effect of a varying average baryon density on the properties of galaxies in neighbouring domains.  In Section 3 we consider general constraints on anthropic baryogenesis models. 
In Section 4 we consider the case of Affleck-Dine baryogenesis.  In Section 5 we present our conclusions.

\section{Galaxy densities in domains of varying baryon number} 

In \cite{lindeax} it was assumed that dark matter is due to an axion field, with $\rho_{DM} \propto \phi_{o}^2$, where $\phi_{o}$ is the initial value of the axion field at the onset of axion oscillations. An important feature of the model is the large enhancement of the dark matter density in galaxies when the {\it average} dark matter density in a domain is varied. This is because the dark matter density in galaxies depends not only on the average dark matter density  but also on the time of matter-radiation equality.  At a given temperature $T$, the average dark matter density and so average total density are increased. As a result, matter-radiation equality occurs at a higher temperature. Perturbation growth starts at a higher temperature and perturbations therefore become non-linear and break away from the expansion of the Universe at a higher temperature. The density of dark matter and baryons in galaxies is approximately equal to the mean density at this time, therefore galaxies will have higher densities of both dark matter and baryons. An increase of the average dark matter density by a factor of 10 was shown to increase the dark matter density in galaxies by $O(10^4)$ \cite{lindeax}. 

   In the case of fixed DM density with varying baryon domains, a similar argument applies but the role of baryons and dark matter is exchanged. The total matter density at a given $T$ in this case can be expressed as 
\be{a3}  \rho = \rho_{B} + \rho_{DM} = (1 + f_{B} r_{o}) \rho_{DM\;o}   ~,\ee
where $r_{o} = \rho_{B\;o}/\rho_{DM\;o}$ (we use $\rho_{B\;o}/\rho_{DM\;o} = 1/5$ throughout, where subscript $o$ denotes values in our domain) and $f_{B} = \rho_{B}/\rho_{B\;o}$ is the enhancement of the average baryon number in a given domain relative to that in our domain. 
The total density at a given $T$ is therefore increased by $K$, where
\be{a4} K \equiv \frac{\rho}{\rho_{o}} = \frac{(1 + f_{B} r_{o})}{(1 + r_{o})}   ~.\ee
Since $\rho \propto T^{3}$ and $\rho_{rad} \propto T^{4}$, the temperature at matter-radiation equality is increased by $K$. 
Perturbation growth is proportional to $T$ during matter-domination. 
The density in dark matter when a perturbation breaks away from the expansion, which we define to occur at $T_{*}$, is therefore increased by $K^3$ relative to our domain, while the total mass density is increased by $K^4$. This follows since (i) the dark matter density at a given $T$ is unchanged between domains while the total density is increased by $K$ and (ii) $T_{*}$ is proportional to $K$. (We are assuming the primordial perturbation is the same in all domains.) The baryon density is therefore increased relative to our domain by 
\be{a5} \frac{\rho_{B}(T_{*})}{\rho_{B\;o}(T_{*\;o})} = \left(K^4(1 + r_{o}) - K^3\right) r_{o}^{-1}    ~.\ee

\begin{figure}[htbp]
\begin{center}
\epsfig{file=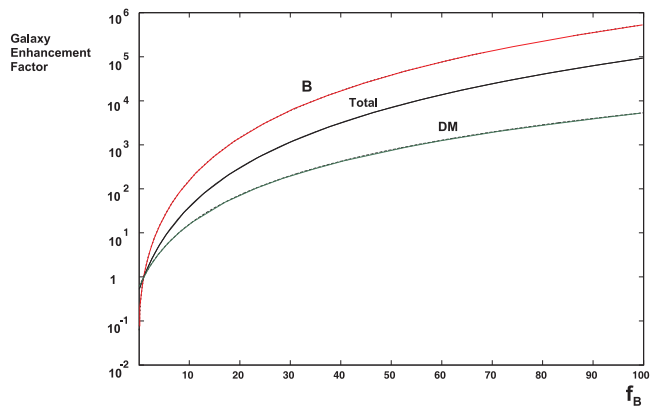, width=0.3\textwidth, angle = -90}
\caption{The enhancement factors of the baryon, dark matter and total mass densities in a galaxy as a function of the enhancement  of the 
average baryon density in a domain, $f_{B}$.}
\label{fig1}
\end{center}
\end{figure}

In Figure 1 we show the enhancement factors for the baryon, dark matter and total mass density in a galaxy as a function of the enhancement factor $f_{B}$ of the average baryon density in a given domain.   
$f_{B} = 10$ (100) will produce an increase in the baryon density in galaxies by a factor of 160 ($5.4 \times 10^5$). (Note that this is a less strong enhancement than in the case where the average dark matter density is increased \cite{lindeax}, simply because the present average baryon density is subdominant.) The strong modification of the properties of galaxies may then provide the necessary anthropic cut-off, since we can expect a strong modification of star formation and the environment around stars in galaxies with much larger baryon and total mass densities than in our domain.

   \section{Baryon isocurvature constraints on anthropic baryogenesis} 

   All particle physics-based baryogenesis or leptogenesis models depend on a CP-violating phase. We will assume in the following that this phase is proportional to the random phase of a complex field $\Phi = \phi e^{i \theta}/\sqrt{2}$, where the phase field is effectively massless during inflation. We will also assume that $n_{B} \propto \theta$ when $\theta \ll 1$. (An explicit example of such a model, Affleck-Dine baryogenesis, will be discussed in the next section.) 

      We expect that all values of $\theta$ will be equally 
probable in the absence of anthropic selection effects. In this case, since the probability of $\theta$ having a value in the range 0.3-3 is approximately 10 times the probability of it being in the range 0.03-0.3, we can say that $\theta \sim 1$ is ten times more likely that $\theta \sim 0.1$. We then assume that the increase in $\theta$ relative to our domain will be anthropically disfavoured by the increase in the baryon density in galaxies, such that we exist in a domain which has close to the maximum probability for observers to exist.  

     For this to be true we require that $\theta$ in our domain is sufficiently small compared with its typical value $\theta \sim \pi$. We will conservatively require that $\theta \lae \theta_{anthropic} = 0.01 \pi$ in order for a domain to be anthropically selected. There are two reasons for this. Firstly, it is unlikely that a domain with such a small value of $\theta$ would be selected at random in the absence of anthropic selection. Secondly, this value of $\theta$ is small enough that there is a significant increase in the baryon density on going from $\theta = 0.01 \pi$ to $\theta \sim \pi$. 
In typical models, the baryon number will be proportional to $\sin(n \theta)/n$ for some integer $n$, such that $n_{B} \propto \theta$ at small $\theta$. In the case of Affleck-Dine baryogenesis,  discussed in the next section, $n = 2$. We will adopt this value throughout. We expect that $\sin(2 \theta)/2$ is approximately $1/4$ in a typical domain. Therefore the ratio of the baryon number in a typical domain to that in our domain, assumed to have $\theta \ll 1$, is approximately $1/(4\theta)$. If $\theta < \theta_{anthropic}$ in our domain, then the average baryon number in a typical domain is at least 8 times larger than in our domain, corresponding to a baryon density in galaxies 80 times larger than in our domain (Figure 1). This is large enough for typical domains to be plausibly anthropically disfavoured relative to the domain we find ourselves in. In contrast, if $\theta > \theta_{anthropic}$, then anthropic selection is less effective, as the baryon density in galaxies cannot be significantly enhanced in a typical domain relative to that in ours. For example, if $\theta = 0.1 \pi$ in our domain then the ratio of the baryon density in a typical domain to our domain is $1/(2 \sin(2 \theta)) = 1.6$. The corresponding enhancement of the baryon density in galaxies is then only by a factor 2.1. This seems unlikely to have a strong effect on the probability of life forming. 
Moreover, domains with $\theta = 0.1 \pi$ are not strongly disfavoured on probabilistic grounds, having approximately 10$\%$ probability relative to more typical domains. Therefore such a domain is not really atypical. This amounts to requiring a 
coincidence between the baryon density in a typical domain and the observed dark matter density. 
Thus $\theta \lae \theta_{anthropic}$ is a reasonable condition for our domain to be plausibly anthropically selected and so for the baryon-to-dark matter ratio to be explained without coincidence.     

     We next discuss how an effectively massless phase field $\theta$ can generate domains of different baryon number. There are two ways this can be done. In both cases we assume that the total number of e-foldings of inflation, $N_{total}$, is much larger than 60. 
(i) At the onset of inflation, we expect the effectively massless angular field to take random values in each initial horizon volume. These volumes are then inflated at $N = 60$ by a factor $e^{N_{total} - 60}$.  In each of these superhorizon-sized domains the phase can take different random values, with $\theta$ essentially constant on the scale of the horizon at $N = 60$. Therefore, in any given horizon volume at $N = 60$, we expect $\theta$ to have a constant but random value. (ii) Even if $\theta$ took only one value throughout the Universe initially, quantum fluctuations of the angular field mean that the value of $\theta$ within a given horizon volume at later times will take random values from 0 to 2$\pi$. In a given horizon volume, the quantum modes of a massless scalar field $\sigma$ of wavelength much larger than the horizon may be considered to be a constant classical field on scales less than the horizon. Therefore each horizon volume has an effectively constant value of $\sigma$. 
In one e-folding, the stretching of quantum modes beyond the horizon will change the mean classical field in each horizon volume by $\Delta \sigma \approx \pm H_{I}/2 \pi$, where $H_{I}$ is the expansion rate during inflation. Therefore the classical field in a given 
horizon volume experiences a random walk. After $\Delta N$ e-foldings, the r.m.s. value of the field in a given horizon volume is $\overline{\sigma} \approx \sqrt{\Delta N} H_{I}/2\pi$ \footnote{This can be formally understood from the power spectrum of fluctuations of the scalar field, ${\cal P}_{\delta \sigma} = H_{I}^{2}/4 \pi^2$. The contribution to $<\delta \sigma^{2}>$ from modes produced over $\Delta N$ e-foldings is then $\Delta N H_{I}^{2}/4 \pi^{2}$.}. For an angular field $\theta = \sigma/\Lambda$, where $\Lambda$ is a constant, this corresponds to a r.m.s. phase $\overline\theta = \overline{\sigma}/\Lambda \approx \sqrt{\Delta N} H_{I}/2\pi\Lambda$. Therefore the r.m.s. phase $\overline{\theta}$ will 
reach $\pi$ after $\Delta N_{\pi} \approx 4 \pi^4 \Lambda^2/H_{I}^2$ e-foldings. Soon after this, the phase will become completely randomized, with all values of $\theta$ equally probable. Therefore, provided that $N_{total} - 60 > \Delta N_{\pi}$,   
a given horizon volume at $N = 60$ will have a random value of $\theta$. In this case, the quantum fluctuations of the angular field which exit the horizon at $N \gg 60$ will produce an essentially constant value of $\theta$ on horizon scales at $N = 60$, while the quantum fluctuations which exit at $N \lae 60$ will produce baryon isocurvature perturbations.

      We next derive general constraints on models which can anthropically select the baryon number while remaining consistent with 
baryon isocurvature constraints. To do this we will consider a generic potential for the field $\Phi$. We assume $V(\Phi)$ has a global $U(1)$ 
symmetry, 
\be{ne1}   V(\Phi) = - \frac{\mu^{2}}{2} \phi^{2} + \frac{\lambda \phi^{n}}{\oM^{n-4}}     ~,\ee
where $\phi = \sqrt{2} |\Phi|$. 
The first term spontaneously breaks the global $U(1)$ symmetry which keeps the angular field massless, while the second term represents generic interaction terms with mass scale $\oM$, where $n \geq 4$.     
(We need only consider the leading order interaction term.) We will set $\lambda = 1$ except when $n = 4$.  
  
    We denote the value of $\phi$ during inflation by $\Lambda$, assumed to be constant. The CP-conserving direction is defined to be $\theta = 0$. Let $\Phi = (\phi_1 + i \phi_2)/\sqrt{2}$, with $\phi_1$ in the $\theta = 0$ direction. Then for $\theta \ll 1$
 \be{e1} \Phi = \frac{1}{\sqrt{2}}(\phi_1 + i \phi_2 ) \approx \frac{1}{\sqrt{2}} \phi (1 + i \theta)    ~\ee
and so\footnote{In the case where $\Phi$ is effectively massless until the onset of baryogenesis, quantum fluctuations of $\phi_{1}$ could also produce baryon isocurvature perturbations. However, in this case 
$\delta \theta = \phi_{2}/\phi_{1} \times \delta \phi_{1}/\phi_{1} = \theta \delta \phi_{1}/\Lambda$. With $\delta \phi_{1} \sim  \delta \phi_{2}$, this contribution to $\delta \theta$ is suppressed by a factor $\theta \ll 1$ compared with that from $\delta \phi_{2}$.}
\be{e2} \delta \theta \approx \frac{\delta \phi_{2}}{\Lambda}   ~.\ee
The baryon isocurvature perturbation due to CP phase fluctuations is then 
\be{e3} S_{B}   = \frac{\delta n_{B}}{n_{B}} = \frac{\delta \theta}{\theta} =  \frac{\delta \phi_{2}}{\theta \Lambda}   ~,\ee
since $n_{B} \propto \theta$.  
Quantum fluctuations of $\phi_{2}$ will produce fluctuations of
$\theta$ and so isocurvature perturbations of baryon number.
The power spectrum of the baryon isocurvature perturbation is therefore  
\be{e5} {\cal P}_{S} = \frac{1}{\theta^2 \Lambda^2} \cP_{\delta \phi_{2}} ~,\ee
where $\cP_{\delta \phi_{2}}$ is the power spectrum of the quantum fluctuations of $\phi_{2}$.  
The power spectrum for a massless field is ${\cal P}_{\delta \phi_{2}} = H_{I}^{2}/4 \pi^2$, therefore
\be{e6} {\cal P}_{S} = \frac{H_{I}^2}{4 \pi^2 \theta^2 \Lambda^2}    ~.\ee
The isocurvature perturbation is parameterized by $\alpha$ \cite{alpha}, where
\be{e7}  \alpha = \left(\frac{\Omega_{B}}{\Omega_{DM}}\right)^{2} 
\frac{\cP_{S}}{\cP_{R}} = \left(\frac{\Omega_{B}}{\Omega_{DM}}\right)^{2}
\frac{H_{I}^{2}}{4 \pi^2 \Lambda^2 \theta^2 \cP_{R}} ~\ee
and $\cP_{R} = (4.8 \times 10^{-5})^2$ is the power spectrum of the curvature perturbation. 
The present WMAP7 upper bound on $\alpha$ (95 $\%$ c.l.) is $\alpha_{lim} = 0.068$ \cite{wmap} \footnote{This is the strongest WMAP7 bound on uncorrelated isocurvature perturbations. $\alpha$ in \cite{wmap} is defined differently from $\alpha$ used here; they are related by 
$\alpha = \alpha_{WMAP}/(1-\alpha_{WMAP})$, with $\alpha_{WMAP} < 0.064$ (95$\%$ c.l.) \cite{wmap}.}.
 \eq{e7} then imposes a baryon isocurvature lower bound, $\theta_{iso}$, on the value of $\theta$  
\be{e9}  \theta > \theta_{iso} = \frac{H_{I}}{\Lambda} \left(2 \pi \alpha_{lim}^{1/2} \cP_{R}^{1/2}\left(\frac{\Omega_{DM}}{\Omega_{B}}\right)\right)^{-1}     ~.\ee
Therefore
\be{e10} \theta > \theta_{iso}  =  2.6 \times 10^{3} \left(\frac{0.068}{\alpha_{lim}}\right)^{1/2}\frac{H_{I}}{\Lambda}    ~.\ee

    In order to have domains which are consistent with both anthropic selection and baryon isocurvature constraints, we require that 
$\theta_{anthropic} \gae \theta > \theta_{iso}$. More realistically, we should require a reasonably wide range of $\theta$ between the upper and lower bounds, otherwise we would require a coincidence between the value of $\theta$ determined by anthropic selection and the narrow range of allowed values. We will conservatively require at least a one order of magnitude difference between $\theta_{iso}$ and $\theta_{anthropic}$. This imposes an upper bound on $H_{I}/\Lambda$, 
\be{e9a} \frac{H_{I}}{\Lambda}  \lae 1.2 \times 10^{-5} \left(\frac{\theta_{iso}}{\theta_{anthropic}}\right) \left(\frac{\alpha_{lim}}{0.068}\right)^{1/2}    ~\ee
This baryon isocurvature bound imposes a strong constraint on anthropic baryogenesis via CP phase fluctuations. To see this we need to consider the value of $\Lambda$ from \eq{ne1}. There are three cases of interest: (i) $\mu \ll H_{I}$, (ii) $\mu \approx H_{I}$ and (iii) $\mu \gg H_{I}$. 

\vspace{0.2cm} 

\noindent \underline{(i)  $\mu \ll H_{I}$:} In this case there is an upper bound on $\Lambda$ from the requirement that $V^{''}(\phi) < H_{I}^{2}$, since we do not expect $\phi$ to be rapidly rolling at $N = 60$ if the total number of e-foldings of inflation is much larger than 60. For $n > 4$ this requires that
\be{e15a}  \Lambda < \left(\frac{H_{I}^{2} \oM^{n-4}}{n \left(n-1\right)}\right)^{\frac{1}{n-2}} ~.\ee  
(We have set $\lambda = 1$ here.) Combining \eq{e15a} with \eq{e9a} gives the upper bound on $H_{I}$ for which it is possible to satisfy both the slow-rolling condition and the baryon isocurvature constraint,
\be{e16} H_{I} \lae \left(1.2 \times 10^{-5} \left(\frac{\theta_{iso}}{\theta_{anthropic}}\right)\left(\frac{\alpha_{lim}}{0.068}\right)^{1/2}    
        \right)^{\frac{n-2}{n-4}} 
\frac{\oM}{\left(n\left(n - 1\right)\right)^{\frac{1}{n - 4}}}    ~.\ee 
In Table 1 we show the upper bound on $H_{I}$ as a function of $n$ for the cases where $\oM = M_{p}$ and $\oM = 10 M_{p}$ when $\theta_{iso}/\theta_{anthropic} = 0.1$. 
\begin{table}[h]
\begin{center}
\begin{tabular}{|c|c|c|}
\hline $n$	&	$H_{I \; max}(\oM = M_{p})$  & $H_{I\;max}(\oM = 10 M_{p} \GeV)$ \\
\hline	$5$	&	$0.21 \GeV$	&	$2.1 \GeV$\\
\hline	$6$	&	$6.3 \times 10^{5} \GeV$	&	$6.3 \times 10^{6} \GeV$\\
\hline	$7$	&	$9.4 \times 10^{7} \GeV$	&	$9.4 \times 10^{8} \GeV$\\
\hline	$8$	&	$1.1 \times 10^{9} \GeV$	&	$1.1 \times 10^{10} \GeV$\\
\hline	$9$	&	$5.4 \times 10^{9} \GeV$	&	$5.4 \times 10^{10} \GeV$\\
\hline	$10$	&	$1.4 \times 10^{10} \GeV$	&	$1.4 \times 10^{11} \GeV$\\
\hline	$12$	&	$5.2 \times 10^{10} \GeV$	&	$5.2 \times 10^{11} \GeV$\\
\hline	$14$	&	$1.1 \times 10^{11} \GeV$	&	$1.1 \times 10^{12} \GeV$\\
\hline	$16$	&	$1.9 \times 10^{11} \GeV$	&	$1.9 \times 10^{12} \GeV$\\
\hline     
 \end{tabular} 
 \caption{\footnotesize{The maximum value of $H_{I}$ for which baryon isocurvature perturbations are sufficiently small when potential lifting terms satisfy the slow-rolling condition.}}  
 \end{center}
 \end{table}

When $n = 4$, the slow-rolling condition becomes 
\be{s1} \Lambda < \frac{H_{I}}{\sqrt{12 \lambda}}    ~.\ee
Combining this with \eq{e9a} gives a baryon isocurvature constraint on $\lambda$, 
\be{s2}   \lambda \lae 1.2 \times 10^{-11}  \left(\frac{\theta_{iso}}{\theta_{anthropic}}\right)^{2} \left(\frac{\alpha_{lim}}{0.068}\right)  ~.\ee

Therefore $\lambda$ must be very highly suppressed to be consistent with the baryon isocurvature constraint when $n = 4$. In addition, the $n = 5$ term in \eq{ne1} must be eliminated unless $H_{I}$ is exceptionally small, $H_{I} \lae 1 \GeV$. In the case of a global $U(1)$ symmetry, the leading order non-renormalizable term is expected to be $n = 6$. This is only compatible with $\oM \leq 10M_{p}$ if $H_{I} < 10^{7} \GeV$. 

   Therefore, in order to generate baryon domains without large isocurvature perturbations when $\mu \ll H_{I}$, either an unusually small value of $H_{I}$ is necessary, requiring a low-scale inflation model, or a non-trivial suppression of non-renormalizable lifting terms is necessary, requiring a more complicated symmetry than would be expected {\it a priori}. 

\vspace{0.2cm} 

\noindent \underline{(ii)  $\mu \approx H_{I}$:} In this case the constraints are very similar to case (i), since the value of $\phi$ at the minimum of \eq{ne1}, which gives $\Lambda$,  is similar to the upper bound on  $\phi$ from the constraint $V^{''} < H_{I}^{2}$, 
\be{ne2}  \Lambda \equiv \phi_{min} =  \left(\frac{\mu^{2} \oM^{n-4}}{n} \right)^{\frac{1}{n-2}}    ~.\ee      
Comparing with \eq{e15a}, we see that $H_{I}^2 \rightarrow \mu^2$ and $n(n-1) \rightarrow n$. Therefore, when $\mu \approx H_{I}$, essentially the same conclusions apply as in case (i).

\noindent \underline{(iii)  $\mu \gg H_{I}$:} In this case the constraints on $n$ are weakened relative to the cases with $\mu \lae H_{I}$. A case of particular interest is that where the renormalizable $n = 4$ term is unsuppressed, as expected in the absence of non-trivial symmetries. In this case $\Lambda$ is given by 
\be{ne3}    \Lambda \equiv \phi_{min} = \frac{\mu}{2 \lambda^{1/2}}   ~.\ee    
The isocurvature constraint \eq{e9a} then implies that 
\be{ne4}  \mu \gae  1.7 \times 10^{15} \GeV \lambda^{1/2} \left(\frac{H_{I}}{10^{10} \GeV}\right) 
\left(\frac{\theta_{anthropic}}{\theta_{iso}}\right) \left(\frac{0.068}{\alpha_{lim}}\right)^{1/2}    ~.\ee
Therefore it is possible to have anthropic baryogenesis consistent with baryon isocurvature constraints for values of $H_{I}$ typical of inflation models ($H_{I} \gae 10^{10} \GeV$)  when $\theta_{iso}/\theta_{anthropic} \leq 0.1$ if $\mu \gae 10^{16} \GeV$. Larger values of $\mu$ permit wider ranges of $\theta$, for example $\mu = M_{p}$ allows $1 \times 10 ^{-5} \lae \theta \lae 0.03$.

   We conclude that it is difficult to generate baryon number domains via fluctuations of the CP violating phase during inflation if $\mu \lae H_{I}$. This case is of particular interest for SUSY models, since we do not expect symmetry-breaking mass squared terms to be larger than $H_{I}$ in that case. For typical inflation models with $H_{I} \gae 10^{10} \GeV$, the baryon isocurvature perturbation is larger than observational limit unless the potential of the field responsible for the CP phase is very flat, requiring suppression of Planck-suppressed non-renormalizable terms to a high order.  However,  it is possible to have anthropic baryogenesis for a generic potential with unsuppressed interaction terms if the 
symmetry-breaking mass term $\mu$ ($ \gg H_{I}$) is sufficiently large. 

     So far we have considered the case where baryon number domains are due to variations of the CP-violating phase of a baryogenesis model, which is assumed to be proportional to the effectively massless phase field of a complex field during inflation. It is also possible that the strength of B-violation could be due to the magnitude of a massless scalar field $\phi$. In this case we would expect the baryon number to be proportional to $\phi^{\gamma}$ for some power $\gamma$. The baryon isocurvature perturbation in this case is $S_{B} = \gamma \delta \phi/\phi$. The baryon number domains are then determined by random values of $\phi$ generated in the same way as the random values of $\theta$. There is an upper bound on $\phi$, $\phi_{max}$, from the condition that $V^{''}(\phi) < H^{2}$ during inflation. To have an anthropically selected baryon density, we require that $\phi/\phi_{max} \ll 1$ in our domain. $S_{B}$ can then be written as $S_{B} = \gamma \;\delta \phi/((\phi/\phi_{max})\phi_{max})$. This is equivalent to $S_{B}$ for the case of varying $\theta$ with $\theta \rightarrow \phi/\phi_{max}$, $\Lambda \rightarrow \phi_{max}$ and $S_{B} \rightarrow \gamma S_{B}$. The baryon isocurvature constraints will therefore be strengthened relative to the case of varying $\theta$ when $\gamma > 1$. As before, in order to have small enough baryon isocurvature perturbations, Planck-suppressed non-renormalizable lifting terms in the $\phi$ potential must be suppressed to a high order in typical inflation models.

   In the next section we will show that that a class of SUSY baryogenesis model, namely Affleck-Dine baryogenesis, can naturally have a sufficiently flat potential to allow domains with $\theta \ll 1$ without violating the baryon isocurvature constraint.

\section{Anthropic Affleck-Dine Baryogenesis}

      The general scenario in which an effectively massless scalar field generates baryon domains has a natural realization in the context of Affleck-Dine baryogenesis \cite{ad}. AD baryogenesis is based on the evolution of a flat direction scalar field of the MSSM. 
For a flat direction of dimension $d$ and UV cut-off approximately $M_{p}$, the superpotential is 
\be{e17}   W = \frac{\lambda \Phi^d}{d! M_{p}^{d-3}}   ~,\ee
where $\Phi$ is the flat direction superfield.
This corresponds to the case where the strength of the physical interaction is characterized by the Planck mass when $\lambda \sim 1$, with the factorial term correctly normalizing the vertex from \eq{e17}. The corresponding scalar potential, including soft-SUSY breaking terms and Hubble corrections, is \cite{drt}  
\be{e18} V(\Phi) = (m^2 - cH^2) |\Phi|^2 + \frac{\lambda^2 |\Phi|^{2(d-1)}}{(d-1)!^2 M_{p}^{2(d-3)} } + (AW + h.c.) 
~,\ee 
where $m$ is the SUSY-breaking scalar mass. $c$ is required to be positive in order to have $\Phi \neq 0$ at early times. After inflation $c$ is generally of order 1. During inflation
$c \sim 1$ for models where inflation is driven by an F-term potential and $c = 0$ for models driven by a D-term potential. As discussed in the previous section, the isocurvature constraint is essentially the same in both cases.  We assume the A-term has no order $H$ correction, therefore $|A| \sim m$. This is easily achieved via a discrete symmetry acting on the inflaton \cite{drt}. The assumption that the A-term is suppressed throughout is essential in order to have an effectively massless angular field until the onset of baryogenesis.

    Oscillations of the AD scalar about $\Phi = 0$ begin once the expansion rate $H$ is equal to $H_{osc} \approx m/c^{1/2}$. (We will set $c = 1$ in the following for simplicity.) The initial amplitude of oscillation is 
\be{e19} |\Phi|_{osc}^2 \approx \kappa_{d} \left(m^2 M_{p}^{2(d-3)}\right)^{1/(d-2)}    ~,\ee
where 
\be{e20} \kappa_{d} = \left( \frac{\left(d-1\right)!^{2}}{\lambda^2 \left(d-1\right)} 
\right)^{1/(d-2)}    ~.\ee
In this we have neglected the A-term, which is of a similar magnitude to the other terms in the potential and so will alter $|\Phi|_{osc}$ only by an O(1) factor.

    The baryon asymmetry is generated by the effect of the B-violating A-term on the evolution of the scalar field. The A-term is comparable to the mass term in the potential when $H \sim m$ at the onset of oscillations. If the initial phase of the A-term is such that $\Phi$ is not aligned with the CP-conserving direction (given by $\theta = 0$ if we assume $A$ and $\lambda$ are real), then the A-term will cause a phase difference between the late-time $\phi_{1}$ and $\phi_{2}$ oscillations, resulting in an elliptical trajectory in the complex $a^{3/2}\Phi$ plane. The effective mass squared splitting between the real and imaginary directions is of order $m^2$ at the onset of oscillations, so the magnitude of the phase difference $\delta$ is typically of the order of 1. 

      The late-time trajectory (when $H \ll H_{osc}$) can be parameterized as 
\be{e21}  \phi_{1} =  \phi(t) \cos(\theta) \sin(mt)   ~\ee
and
\be{e22}  \phi_{2} =  \phi(t) \sin(\theta) \sin(mt + \delta)   ~,\ee
where $\phi(t) \propto a^{-3/2}$ and $\theta$ is the initial phase of $\Phi$ relative to the CP-conserving direction, which is the $\phi_{1}$ direction here. The baryon asymmetry is then 
\be{e23} n_{B} = i B(\Phi) \left( \dot{\Phi}^{\dagger} \Phi - \Phi^{\dagger} \dot{\Phi} \right) 
= B(\Phi) m \phi^{2}(t) \sin(2 \theta) \sin(-\delta)  ~,\ee
where $B(\Phi)$ is the baryon number of $\Phi$. 
The angle $\theta$ will then determine the baryon asymmetry. 
For an initial phase angle close to the CP conserving direction ($\theta \ll 1$), as required for an atypical domain, the baryon asymmetry is therefore
\be{e24} n_{B} \approx 2 m B(\Phi) \phi^{2}(t) \theta \sin(-\delta)  ~.\ee
The value of $\phi(t)$ can be estimated from the initial value at the onset of oscillations and the assumption that $\phi \propto a^{-3/2}$ once $H < H_{osc}$, in which case $\phi(t) \approx \sqrt{2} 
|\Phi_{osc}| (a_{osc}/a)^{3/2}$. The baryon asymmetry at present is therefore
\be{e25} \eta_{B} \approx  \frac{\kappa_{d} B(\Phi) T_{R}}{M_{p}} \left(\frac{M_{p}}{m}\right)^{\frac{d-4}{d-2}} \theta \sin(-\delta)  ~.\ee      
This determines the reheating temperature necessary to generate the observed baryon number ($\eta_{B\;obs} = 1.5 \times 10^{-10}$),  
\be{e26} T_{R} \approx \frac{\eta_{B\;obs} M_{p}}{|B(\Phi)| \kappa_{d}} \left( \frac{m}{M_{p}} \right)^{\frac{d-4}{d-2}} \frac{1}{\sin(|\delta|) \theta}  
~.\ee
In the following we will refer to AD baryogenesis in a domain with $\theta \lae \theta_{anthropic} \ll 1$ as Anthropic Affleck-Dine Baryogenesis (AADB).
Note that since $\theta$ is small compared with $\pi$ in the case of AADB, the reheating temperature can be much larger than in conventional AD baryogenesis.

When $c \sim 1$, as in F-term inflation models, the value of $\Lambda$ is fixed by the minimum of the potential during inflation, which is determined by the $-c H_{I}^{2} |\Phi|^2$ term and non-renormalizable term in \eq{e18}, which gives
 \be{e27} \Lambda \equiv \sqrt{2} |\Phi|_{min} \approx \sqrt{2} \kappa_{d}^{1/2} \left(|c| H_{I}^2 M_{p}^{2(d-3)}\right)^{1/2(d-2)}    ~.\ee 
The baryon isocurvature perturbation from \eq{e27} and \eq{e7} is then 
\be{e28}  \alpha = \left(\frac{\Omega_{B}}{\Omega_{DM}}\right)^{2} \frac{1}{\kappa_{d}} 
\frac{1}{8 \pi^2 \theta^2 \cP_{R}} \left( \frac{H_{I}}{M_{p}} \right)^{2 \left( \frac{d-3}{d-2} \right)}
~.\ee

   As a specific example we will first consider a $d = 6$ flat direction of the form $W \propto \left(u^{c}d^{c}d^{c}\right)^2$. In this case $B(\Phi) = 1/3$. The observed baryon asymmetry then fixes the reheating temperature to be 
\be{e32} T_{R} \approx 96 \GeV \sqrt{\lambda} \left( \frac{\theta_{anthropic}}{\theta} \right)
\left(\frac{m}{1 \TeV}\right)^{1/2} \frac{1}{\sin (\left|\delta\right|)}  ~.\ee 
This is well above the freeze-out temperature $T_{\chi}$ of neutralino LSPs $\chi$ when the LSP mass is less than a few TeV ($T_{\chi} \approx m_{\chi}/20$), therefore the model is consistent with thermal relic neutralino dark matter.  
The isocurvature perturbation is\footnote{Baryon isocurvature perturbations in AD baryogenesis were first discussed in \cite{kmiso} and later in \cite{jmiso} and \cite{jiso}.}  
\be{e29} \alpha = \frac{1}{\kappa_{6} \theta^2} \left(\frac{\Omega_{B}}{\Omega_{DM}}\right)^{2}
\frac{1}{8 \pi^2 \cP_{R}} \left(\frac{H_{I}}{M_{p}}\right)^{3/2}   ~,\ee
therefore
\be{e30}  \alpha = 0.27 \sqrt{\lambda} \left( \frac{\theta_{anthropic}}{\theta}\right)^2 \left(\frac{H_{I}}{10^{13} \GeV}\right)^{3/2}   ~.\ee

  As an example of a realistic SUSY inflation model, we will consider the case of F-term hybrid inflation.  The superpotential of F-term hybrid inflation \cite{fti,lr} is $\kappa S (\mu^2 - \Phi_{+} \Phi_{-})$, where $\Phi_{+},\Phi_{-}$ are oppositely charged superfields. The observed curvature perturbation fixes $\mu= 5.6 \times 10^{15} \GeV$ and so $H_{I} = 7.4 \times 10^{12} \kappa \GeV$. This is true as long as $|S|^2$ is large compared $|S|_{c}^2$. where $|S|_{c} = \mu$ is the value at which the phase transition ending inflation occurs. Since during inflation $|S| = \kappa \sqrt{N} M_{p} /2 \sqrt{2} \pi$, this condition is satisfied at $N = 60$ if $\kappa$ is significantly larger than $0.0026$. We will use $\kappa = 0.005$ as a lower bound for conventional F-term inflation to be valid.

  In this case \eq{e30} gives
\be{e31}  \alpha = 6.1 \times 10^{-5} \sqrt{\lambda} \left( \frac{\theta_{anthropic}}{\theta}\right)^2 \left(\frac{\kappa}{0.005}\right)^{3/2} 
  ~.\ee 
This can also be expressed in terms of the lower bound on $\theta$, $\theta_{iso}$, for which $\alpha < \alpha_{lim}$, 
\be{e31a} \theta > \theta_{iso} = 9.4 \times 10^{-4} \left(\frac{0.068}{\alpha_{lim}}\right)^{1/2} \lambda^{1/4} \left(\frac{\kappa}{0.005}\right)^{3/4} ~.\ee
Thus in $d = 6$ AD baryogenesis, in the context of a conventional F-term hybrid inflation model, it is possible to have an atypical domain with $\theta \ll 1$ which is consistent with the baryon isocurvature constraint \eq{e31a}.

    The lower bound on $\theta$ from \eq{e31a} is approximately 30 times smaller than $\theta_{anthropic}$ when $\kappa = 0.005$, therefore a reasonable range of $\theta$ can satisfy $\theta_{iso} < \theta \lae \theta_{anthropic}$.  However, the range of $\theta$ is narrow enough that the value of $\theta$ in our domain could easily be close to the isocurvature bound. (There is no reason for $\theta$ to be close to $\theta_{anthropic}$.) Therefore it is possible that baryon isocurvature perturbations in this model will be large enough to be observed in the future. 

 The reason AD baryogenesis can be compatible with isocurvature constraints is the efficient suppression of non-renormalizable terms in the flat-direction potential via a combination of SUSY, R-parity and SM gauge symmetries. This serves as a sufficiently complex symmetry to maintain the flatness of the potential to a high order.

      The range of $\theta$ can be increased by considering flat directions with larger $d$. However, in this case the reheating temperature is below the freeze-out temperature of neutralino dark matter for most or all of the allowed $\theta$ range, ruling out thermal relic neutralino dark matter. 
For a $d = 8$ flat direction the reheating temperature and $\theta$ lower bound are 
\be{e40} T_{R} \approx 0.15 \GeV \lambda^{1/3} \left( \frac{\theta_{anthropic}}{\theta} \right)
\left(\frac{m}{1 \TeV}\right)^{2/3} \frac{1}{\sin (\left|\delta\right|)}  ~\ee 
and 
\be{e41} \theta > \theta_{iso} = 1.6 \times 10^{-4} \left(\frac{0.068}{\alpha_{lim}}\right)^{1/2} \lambda^{1/6} \left(\frac{\kappa}{0.005}\right)^{5/6} ~.\ee
For a $d = 10$ flat direction these become
\be{e42} T_{R} \approx 5.4 \times 10^{-3} \GeV \lambda^{1/4} \left( \frac{\theta_{anthropic}}{\theta} \right)
\left(\frac{m}{1 \TeV}\right)^{2/3} \frac{1}{\sin (\left|\delta\right|)}  ~\ee 
and 
\be{e43} \theta > \theta_{iso} = 6.0 \times 10^{-5} \left(\frac{0.068}{\alpha_{lim}}\right)^{1/2} \lambda^{1/8} \left(\frac{\kappa}{0.005}\right)^{7/8} ~.\ee
Therefore for $d = 8$ flat directions there is a factor of 200 between $\theta_{iso}$ and $\theta_{anthropic}$. However, $T_{R}$ can be larger than the neutralino LSP freeze-out temperature only if $\theta$ is close to $\theta_{iso}$. For $d =10$ flat direction there is a factor of 500 between $\theta_{iso}$ and $\theta_{anthropic}$, but the reheating temperature is at most a few GeV, well below the freeze-out temperature of neutralino LSPs.

\section{Conclusions and Discussion}

        The similarity of the observed baryon and dark matter densities suggests that there is a physical process connecting them. This similarity is particularly difficult to understand in the case of thermal relic WIMP dark matter,  since this requires an explanation of why the baryon abundance is within an order of magnitude of the thermal relic dark matter density, ruling out simple co-production of baryons and dark matter. Here we have considered an  anthropic selection mechanism based on superhorizon-sized domains of varying baryon density.

   We have discussed a general framework, {\it anthropic baryogenesis}, in which the domains are generated by variations of a complex scalar field $\Phi$ and anthropic selection is assumed to disfavour domains in which galaxies have a baryon density which is much larger than in our domain. 

   Baryon isocurvature perturbations impose  strong constraints on anthropic baryogenesis. 
In the case where the $\Phi$ mass during inflation satisfies $|m_{\Phi}| \lae H_{I}$,  either an inflation model with an unusually small expansion rate during inflation, $H_{I} < 10^{7} \GeV$, or a high-order suppression of Planck-suppressed terms in the $\Phi$ potential is necessary to suppress the baryon isocurvature perturbation. The need to suppress non-renormalizable terms to a high order rules out models with only simple symmetries. This case is relevant to SUSY models, since in that case the mass-squared terms are at most of order $H^2$. 

     Alternatively, an unsuppressed $\Phi$ potential is possible if the symmetry-breaking mass term $\mu$ in the $\Phi$ potential is sufficiently large, $\mu \gae 10^{16} \GeV$.

   The necessary suppression of potential terms is natural in the case of Affleck-Dine baryogenesis, where the combination of the SM gauge symmetry, SUSY and R-parity provides a sufficiently complex symmetry to suppress the non-renormalizable terms to a high order. We have considered Affleck-Dine baryogenesis for the case of a $d = 6$ $(u^{c}d^{c}d^{c})^2$ flat-direction in the context of F-term hybrid inflation.  With inflaton superpotential coupling $\kappa = 0.005$, the value of the CP-violating phase in our domain must be in the range $0.001 \lae \theta \lae 0.03$,
where the lower bound is the isocurvature constraint and the upper bound is the value below which the baryon density can be considered to be anthropically selected. 
The existence of this range allows $\theta$ in our domain to be small enough for anthropic selection to function but large enough to evade large baryon isocurvature perturbations. 
Since $\theta$ in our domain, which is determined anthropically, can take any value within this range, it is possible that baryon isocurvature perturbations will be large enough to be observed in the future.

  In our model we have considered all the parameters of the Universe to be fixed to their observed values except the baryon density. In particular, we have considered the dark matter density to be fixed and equal to its value in the observed Universe. This raises an important issue for the class of anthropic selection model considered here. The underlying assumption is that there is a critical baryon density above which life is anthropically disfavoured. The baryon density in a domain will take the largest value possible up to anthropic selection effects, therefore the baryon density will be close to this critical density. 
It is therefore assumed that this critical baryon density is close to the observed baryon density. But this does not explain why the observed dark matter density, which is assumed to be a fixed parameter,  is also close to the critical baryon density. (A similar problem arises in the model of \cite{lindeax}, where the baryon number is assumed to be fixed and the axion dark matter density varies between domains. In this case it is not explained why the fixed baryon density is close to the critical density.) In order to achieve a complete solution, it may be necessary for both the baryon and dark matter densities to vary independently between domains. In this case the baryon density in a domain will have the highest probability when it is close to the critical density. The dark matter density will then have the highest probability when it is close to baryon density, since there will be a rapid increase in the baryon and dark matter densities in galaxies once $\Omega_{DM} > \Omega_{B}$ \cite{lindeax}. In the case of thermal relic WIMP dark matter, this suggests that a domain-dependent Higgs expectation value and so domain-dependent weak scale is necessary. Alternatively, axion dark matter combined with an anthropic baryogenesis model, such as Affleck-Dine baryogenesis, could provide the basis for such a model. We will return to this possibility in future work.

\section*{Acknowledgements}
The work of JM is supported by the Lancaster-Manchester-Sheffield Consortium for Fundamental Physics under STFC grant
ST/J000418/1.


\end{document}